\documentstyle[12pt]{article}

\begin{document}
\newcommand {\be}{\begin{equation}}
\newcommand {\ee}{\end{equation}}
\newcommand {\bea}{\begin{array}}
\newcommand {\cl}{\centerline}
\newcommand {\eea}{\end{array}}
\renewcommand {\theequation}{\thesection.\arabic{equation}}
\newcommand {\newsection}{\setcounter{equation}{0}\section}

\baselineskip 0.65 cm
\begin{flushright}
IPM/P-98/19 \\
hep-th/9810072
\end{flushright}
\begin{center}
 {\Large {\bf  Noncommutative Geometry from Strings and Branes}}
\vskip .5cm

F. Ardalan$^{a,b}$, H. Arfaei$^{a,b}$ and M.M. Sheikh-Jabbari$^a$
\footnote{ E-mail:  ardalan, arfaei, jabbari@theory.ipm.ac.ir } \\

\vskip .5cm

 {\it $^a$ Institute for studies in theoretical Physics and Mathematics 
IPM,

 P.O.Box 19395-5531, Tehran, Iran}\\
{\it and}
\\
{\it $^b$ Department of Physics Sharif University of Technology, 

P.O.Box 11365-9161, Tehran, Iran}
\end{center}

\vskip 2cm
\begin{abstract}
Noncommutative torus compactification of Matrix model is shown to be a 
direct consequence of quantization of the open strings attached to a 
D-membrane with
a non-vanishing background $B$ field. We calculate the BPS spectrum 
of such a brane system using both string theory results and DBI action.
The DBI action leads to a new transformation property of the
compactification radii under the $SL(2,Z)_N$ transformations.

\end{abstract}
\newpage
\section{Introduction}
Recently noncommutativity of space-time coordinates has emerged in a number of 
occasions in string theory. After the discovery of the significance  of D-branes 
as carriers of RR charge in string theory [1], it was observed that embedding 
coordinates of D-branes are in fact noncommutative [2]. The reason for this
surprising result is that, dynamics of N coincident 
$D_{p}$-branes in low energy regime can be shown to be described by a supersymmetric 
Yang-Mills (SYM) SU(N) gauge theory in $p+1$ dimensions, obtained by dimensional 
reduction of 10 dimensional $N=1$ SYM theory. Through this dimensional 
reduction, the components of gauge field corresponding to transverse 
direction to the brane behave as scalar fields of the $(p+1)$ dimensional gauge  
theory. These scalars are the transverse coordinates of the D-brane, and hence
result in the noncommutativity of space coordinates.

These noncommutative coordinates in the case of 0-branes are elevated to the 
dynamical variables of Matrix-model which is conjectured to describe the strong 
coupling limit of string theory, or M-theory, in the infinite momentum frame [3].

In Matrix-model the noncommutativity of matrices, and therefore the 
coordinates, become significant in substringy scales, as expected from general 
quantum gravitational considerations.

Another type of space-time noncommutativity has been recently observed in 
M-theory which is naively different from the above noncommutativity.
It arose from the application of the non-commutative geometry (NCG) techniques
pioneered by A. Connes to the Matrix-model compactifications [4].  

According to Matrix-model conjecture, each momentum sector of the discrete
light cone quantization (DLCQ) of M-theory is described by a maximally 
supersymmetric Matrix-model (or SYM), with the light cone momentum identified  
with the rank of gauge group. This conjecture has passed many consistency
checks; for a review of Matrix model see [6,7,8].
To be a formulation of M-theory, Matrix model must describe string theory 
when compactified on a circle. Moreover, one should consider further 
compactifications of Matrix model, and check the conjectured U-duality groups
of M-theory in various compactifications.
But Matrix model compactifications involve complicated operations and 
it is not at all clear how to obtain them in general. A class of
toroidal compactifications were constructed in early stages of Matrix model 
development, which relied on a certain commutative subalgebra of matrices [8,9]. 
In a certain sense this subalgebra is an equivalent description of the manifold 
of torus on which compactification is performed.

It was observed by  Connes, Douglas and Schwarz (CDS) that generalizing this 
same algebraic description of the manifold of compactification,  in the spirit of 
NCG, to a noncommutative torus, it is   
possible to arrive at a different compactification of Matrix-model and 
different physical consequences, which is equivalent to adding a constant  
3-form background in the 11 dimensional supergravity. A major result is that 
the SYM theory of commutative torus compactification now becomes a  
"deformed" SYM theory, with important non-local interactions introduced [10,11].  
Soon after, it was observed by Douglas and Hull [10] that deformed SYM theory 
and, therefore indirectly, the noncommutative torus (NCT) compactification 
is a natural consequence of certain D-brane configurations in string theory.

Subsequently compactification on more complicated spaces were considered in
[12] and various properties of the deformed SYM theory and their relation to
string theory were studied [13,14,15].

It is then clear that there is a close connection between non-zero constant 
background Kalb-Ramond anti-symmetric field ($B_{\mu\nu}$) and deformation of 
the torus of compactification of Matrix model and the non-locality of the
resultant deformed field theory on torus. Yet it was not obvious how the 
turning the background B field on, causes the coordinates to become noncommutative
and how this noncommutativity differs from that of coincident D-branes.

In this article \footnote{A preliminary version of this work presented in 
PASCOS98 [16]} we propose an explicit construction of this noncommutativity and 
compare it with the noncommutativity due to coincident D-branes. 
We will show that a string theory membrane wrapped around $T^2$ in the presence 
of background B field, manifests noncommutative coordinates as a simple consequence of 
canonical commutation relations. 
We then show that applying T-duality and using the DVV string matrix theory
[17] relation of Matrix model to string theory, results in Matrix model 
compactification on a deformed torus.

The plan of the paper s as following. In section 2, we review the CDS construction
([4]). Section 3, contains the explicit noncommutative coordinate construction  
of the wrapped membrane; and section 4, is devoted to the mass spectrum and
its symmetries. In section 5, we will compare our string theory results with
the other works in this subject and discuss the role of DBI action. 

\section{ Compactification on a noncommutative torus}

Matrix model describes M-theory in the infinite momentum frame. The 
dynamical variables  are
$N\times N$ matrices which are function of time, and $N$ is taken to infinity.
Matrix model is  described by the supersymmetric action,

\be
I={1 \over 2g\sqrt{\alpha'}}\int d\tau \;\;\;Tr \; \biggl\{ \dot{X_a}\dot{X_a}
+{1 \over {(2\pi\alpha')}^2}\sum_{a<b}[X^a,X^b]^2 \\
+{i \over {2\pi\alpha'}} \Psi^T\dot{\Psi}- {1 \over {(2\pi\alpha')}^2} \Psi^T
\Gamma_a[X^a,\Psi]\biggr\}.
\ee
$X^a, a=1,...,9$ are bosonic hermitian matrices and $\Psi$ are 16 component
spinors. $\Gamma^a$ are $SO(9)$ Dirac matrices. Classical time independent solutions
have commuting $X^a$, therefore simultaneously diagonalizable, corresponding to 
the  classical coordinates of $N$ 0-branes. In general off-diagonal elements of
$X^a$ correspond to substringy noncommutative structure of M-theory. This 
theory as a candidate for M-theory has passed a number of tests.

Compactification of coordinate $X$ of Matrix model on a space-like circle of 
radius $R$ has been shown [9] to require existence of the matrix $U$ with the 
property 
\be
\bea{cc}
UXU^{-1}=X+R, \\
UX^a U^{-1}=X^a  \;\;\;\;\;\; X^a\neq X ,\\
U\Psi U^{-1}=\Psi.
\eea
\ee
It was then shown that the solution of these equations can be written in terms of 
a covariant derivative
\be \bea{cc}
X=i{\partial \over \partial \sigma}+ A, \\
U=e^{i\sigma R},
\eea \ee
and when substituted in the original action $I$, it becomes that of a $(1+1)$ 
dimensional SYM on the dual circle. This $(1+1)$ dimensional space 
is parametrized by $(\sigma,\tau)$, and the coupling constant of this theory
is $g^2_{YM} \sim {1 \over R}$.

It was then shown that this $(1+1)$ dimensional SYM theory is identical to the
IIA string theory for string scales, as expected [17]. The DVV map which relates
the matrices to the strings plays an important role in our description of the 
noncommutativity in Matrix theory. The coupling constant of the string varies 
as inverse of the $g_{YM}^2$, $g_s={1 \over \alpha' g_{YM}^2}$, the dimension of
the matrices, $N$, is carried into the light cone energy $p^+$ of strings and the 
eigenvalues of the matrices correspond to $N$ free strings in the limit of vanishing
$g_s$.

Compactification on a 2 torus is similarly accomplished by solving the equations
\be
\bea{cc}
U_1X_1U_1^{-1}=X_1+R_1 \\
U_2X_2U_2^{-1}=X_2+R_2 \\
UX^a U^{-1}=X^a  \;\;\;\;\;\; a\neq 1,2\\
U\Psi U^{-1}=\Psi
\eea
\ee
But now consistency between these equations requires:
\be
U_1U_2=e^{i\theta}U_2U_1,
\ee
for some real number $\theta$; where for the usual commutative torus, 
$\theta=0$. We will later see that in fact a {\it rational} $\theta$ will 
also give a commutative torus. Again it is easily seen that for $\theta=0$,
\be \bea{cc}
X_i=i{\partial \over \partial \sigma_i}+ A_i \;\;\;\; , i=1,2 \\
U_i=e^{i\sigma_i R_i}.
\eea \ee
is a solution of eq. (2.4) and (2.5) and its insertion in the action results 
in the 2+1 dimensional SYM on the dual torus. Here $\sigma_i$ parameterize the 
dual two-torus. 

Connes, Douglas and Schwarz [4] observed that in Eq. (2.5), $\theta$ can be 
taken different from zero and it corresponds to compactification on a 
noncommutative torus (NCT) and the resulting gauge theory is the SYM with the commutator 
of the gauge fields replaced by the Moyal bracket. 
The central idea of NCG is, starting from the equivalence of a manifold with 
the $c^*$ algebra of functions over that manifold, to generalize to a 
noncommutative $c^*$ algebra [18]. Thus, the algebra generated by the commuting 
matrices $U_1$ and $U_2$ in the case of usual $T^2$, is generalized [12,18,19] 
to the algebra generated by $U_1$ and $U_2$ satisfying the relation (2.5), which 
now defines a "noncommutative" torus, $T^2_{\theta}$. 
The solutions of (2.5) are then,
\be
X_i=-iR_i\partial_i +A_i,
\ee
where $A_i$ now are functions of $\tilde{U_i}$, with $\tilde{U_i}$ satisfying 
\be
\bea{cc}
\tilde{U_1}\tilde{U_2}=\;\;\; 
e^{-i\theta}\tilde{U_2}\tilde{U_1}$, $\;\;\;\;U_i\tilde{U_j}=\tilde{U_j}U_i \\  

 [\partial_i, \tilde{U_j}]=i\delta_{ij} \tilde{U_j} ;\;\; i,j=1,2.
\eea
\ee
Substituting them in the action, we get the SYM theory on the NCT dual to the  
original one, with the essential modification being, the replacement of 
commutators of gauge fields by the Moyal bracket,
\be\bea{cc}
\bigl\{ A,B\bigr\}=A*B-B*A ,\\
A*B(\sigma)=e^{-i\theta(\partial'_1\partial''_2-\partial'_2\partial''_1)} A(\sigma')B(\sigma'')|_
{\sigma'=\sigma''=\sigma}.
\eea \ee
with $\sigma=(\sigma_1,\sigma_2)$. Moyal bracket introduces 
non-locality into the theory. This theory obviously suffers from lack of 
Lorentz invariance in the substringy scales; however, has better convergence 
properties compared with the ordinary SYM theory [10,13].

An important test of the conjecture that the compactification of the Matrix
model in the presence of non-zero 3-form background field is equivalent to 
the SYM theory on a NCT, is  comparison of the mass spectra of the two theories. 
This comparison, in the case of BPS states, was carried out in CDS, by 
giving the BPS spectrum of SYM on NCT with the low energy BPS states in 
11 dimensional supergravity in the
presence of the three form $C$ in the light cone direction. 

Ho [19] calculated the same BPS spectrum for the Matrix theory compactified on 
NCT, with certain modifications, i.e. to take into account the longitudinal and 
transverse membrane winding modes, which is in fact equivalent, as we will show 
in section 4, to turning on the $B_{\mu\nu}$ background field. 
He obtained the energy of BPS states 
\footnote{ We would like to thank the referee for pointing out that, 
although Ho obtained this result from a modified Matrix model action which 
had no justification, the spectrum is still valid.},

\be \bea{cc}
E={R \over {n-m\theta}}\biggl\{ {1 \over 2}\bigl( {n_i-m_i\theta\over R_i}\bigr)^2
+{V^2 \over 2}\bigl[ m+(n-m\theta)\gamma\bigr]^2 \\
+2\pi \sqrt{(R_1w_1)^2+(R_2w_2)^2} \biggr\}. 
\eea\ee
where $V=(2\pi)^2R_1R_2$ and ${n_i \over R_i}$ are KK momenta conjugate to 
$X_i$; $m_i=\epsilon_{ij}m_{j-}$, with $m_{i-}$ winding number of the longitudinal 
membrane along $X_i$ and $X_-$ direction; $R$ the compactification radius 
along the $X_-$ direction and $w_i$ are the momenta of BPS states due to the 
transverse coordinates and are constrained by:
\be
w_i=\epsilon_{ij}(nm_j-mn_j).
\ee
Moreover $n$ is the dimension of matrices (number of 0-branes), $m$ the winding number of  
the membrane around torus and $\theta$ is the deformation parameter of the torus.
It is then noted that this spectrum is the same as that obtained from the SYM 
theory on NCT, where $m=\epsilon_{ij}m_{ij}$ [4,19]. 
The term involving $\gamma$ is essentially put in an ad hoc manner and is mainly
needed for the $Sl(2,Z)_N$ symmetry below. This term corresponds to an 
arbitrariness in the mass formula of CDS.
We will summarize and compare these results with CDS's and with ours at the end of 
section 4. 

An important property of the mass spectrum (2.11) is its $SL(2,Z)_N$ invariance
generated by
\be\bea{cc}
\theta \rightarrow {-1 \over \theta} \\
m \rightarrow n \;\;\;\; , \;\;\;\; n \rightarrow -m \\
m_i \rightarrow n_i \;\;\; , \;\;\;\; n_i \rightarrow -m_i \\
\gamma \rightarrow -\theta(\theta\gamma+1) \\
R_i \rightarrow \theta^{-2/3} R_i \;\;\; , \;\;\;\;\; R \rightarrow 
\theta^{-1/3} R
\eea\ee
and
\be\bea{cc}
\theta \rightarrow  \theta+1 \\
n \rightarrow n+m \;\;\;\ , \;\;\;\;\; m \rightarrow m \\
n_i \rightarrow n_i+m_{i} \;\;\;\ , \;\;\;\;\; m_i \rightarrow m_i.
\eea\ee
This invariance is to be expected  on the basis of the NCG considerations. It
is the SL(2,Z) invariance of the $c^*$-algebra defining the NCT [4].

We note that from noncommutative geometric arguments, CDS observed that 
the commutator of the NCT coordinates should satisfy
\be
[X^1,X^2]=2\pi i R_1R_2 {m \over n-m\theta}
\ee
We will later obtain this relation from string theory.

\section{ Noncommutativity from string theory}
\setcounter{equation}{0} 
In this section we trace the noncommutativity of the Matrix model 
compactification in string theory formulated in the presence of the 
antisymmetric background field. The noncommutativity appears in string theory 
when we consider D-branes living in the $B_{\mu\nu}$ background.

We begin with the action of Fundamental strings ending on a D-membrane in the 
background of the antisymmetric field, $B_{\mu\nu}$ [20]:

\be\bea{cc}
S= {1 \over 4\pi\alpha'} \int_{\Sigma} d^2\sigma \bigl[ \eta_{\mu\nu}
\partial_aX^{\mu}\partial_bX^{\nu}g^{ab}+ \epsilon^{ab} B_{\mu\nu}\partial_a
X^{\mu}\partial_bX^{\nu}+ 
{1 \over 2\pi\alpha'}\oint_{\partial \Sigma} d \tau A_i \partial_{\tau}\zeta^i, 
\eea\ee
where $A_i,\ i=0,1,2$ is the $U(1)$ gauge field living on the D-membrane and
$\zeta^i$ its internal coordinates. 
The action is invariant under the combined gauge transformation [2]
\be \bea{cc}
B_{\mu\nu}\rightarrow B_{\mu\nu}+\partial_{\mu}\Lambda_{\nu}-\partial_{\nu}\Lambda_{\mu} \\
A_{\mu} \rightarrow A_{\mu}-\Lambda_{\mu}.
\eea\ee
The gauge invariant field strength is then 
\be
{\cal F}_{\mu\nu}=B_{\mu\nu}-F_{\mu\nu} \;\;\;\;\; ,\;\; 
F_{\mu\nu}=\partial_{[\mu}A_{\nu]}.
\ee
Variation of the action $S$, leads to the following mixed boundary conditions 

\be
\left\{  \begin{array}{cc}
\partial_{\sigma}X^0=0 \\
\partial_{\sigma}X^{1}+{\cal F} \partial_{\tau}X^2=0 \\
\partial_{\sigma}X^{2}- {\cal F} \partial_{\tau}X^{1}=0  \;\;\; , 
{\cal F}={\cal F}_{12}\\
\partial_{\tau} X^a=0 \;\;\;\ ,\;\;\; a=3,...,9.
\end{array}\right.
\ee
Imposing the canonical commutation relation on $X^i$ and its conjugate momenta
$P^i$, $i=1,2$:
\be 
P^1=\partial_{\tau}X^1-{\cal F}\partial_{\sigma}X^2 \;\;\; ,\;\;\; 
P^2=\partial_{\tau}X^2+{\cal F}\partial_{\sigma}X^1.
\ee
\be
[X^{\mu}(\sigma,\tau),P^{\nu}(\sigma',\tau)]=i\eta^{\mu\nu}\delta(\sigma-\sigma').
\ee
Leads to the non-trivial relation 
\footnote{A non-zero ${\cal F}_{0i}$, will not give
any noncommutativity between $X^0$ and $X_i$. This is the effect of the  
world sheet metric signature.}

\be
[X^1(\sigma,\tau),X^2(\sigma',\tau)]=2\pi i {\cal F} \theta(\sigma-\sigma').
\ee

Mode expansions for $X^1$ and $X^2$ consistent with our boundary conditions, 
are
\be
\left\{  \begin{array}{cc}
X^{1}=x_0^1+(p^1 \tau-{\cal F} p^2 \sigma)+ \sum_{n\neq 0} {e^{-in\tau} \over n}
\bigl(ia^1_n \cos n\sigma +{\cal F}  a^2_n \sin n\sigma \bigr) \\
X^2=x_0^2+(p^2 \tau+{\cal F} p^1 \sigma)+ \sum_{n\neq 0} {e^{-in\tau} \over n}
\bigl(ia^2_n \cos n\sigma -{\cal F} a^1_n \sin n\sigma \bigr)
\end{array}\right.
\ee
From which the center of mass coordinates
\be
x^i={1 \over \pi}\int X^i(\sigma,\tau)\ d\sigma
\ee
satisfy
\be
[x^1,x^2]=\pi i{\cal F}.
\ee
We claim that this noncommutativity of space coordinates is at the root of the
geometric noncommutativity which appears in the compactification of Matrix 
model on a torus in the background 3-form field, as described in section 2.
To show  this, we will map the coordinate $X^i$ to the Matrix model variables 
via the string matrix model of DVV, and use it to construct the NCT compactification 
discussed in the previous section.

But, first we would like to elaborate on the connection between the noncommutativity
of (3.10) and the noncommutativity which appears in the transverse 
coordinates of several coincident D-branes. The point is that it has been shown that 
D-membranes with a non-zero U(1) gauge field in the background 
contain a distribution of 0-branes proportional to {\cal F} [16,21,22]. 
In previous works only the zero $B_{\mu\nu}$ case were considered.
These 0-branes as described by the Matrix model, live on a torus with 
a D-membrane wrapped on, with noncommutative coordinates $X^1$ and $X^2$,
\be
[X^1,X^2]=if.
\ee
The proportionality constant $f$ is given by the $U(1)$ gauge field strength
${\cal F}$ [22].

It is quite remarkable that the elaborate mechanism which produces the original 
noncommutativity in the description of D-branes, first discovered by Witten, 
and leads through a set of subtle arguments to the particular form of the 
commutation relation in (3.11), should  be simply
derived from the string action (3.1) in the presence of  
the ${\cal F}$ and mixed boundary conditions for zero $B$ field background.

We will shortly see that the noncommutativity (3.10), under certain circumstances, 
leads to the noncommutative torus compactification of CDS.
To see this we compactify the $X^i$ direction and wrap the 2-brane around the 
2-torus and use the center of mass coordinates $x^i$ and their conjugate momenta 
to construct the generators of the $c^*$ algebra of the noncommutative torus;
proving that the compactification, in the presence of $U(1)$ field strength,
for D-membrane requires a NCT. Thus we demand to solve the compactification 
equations for the membrane coordinates $x^i$,
\be
\bea{cc}
U_1x^1U_1^{-1}=x^1+ R_1 \\
U_2x^2U_2^{-1}=x^2+ R_2 \\
U_ix^j U_i^{-1}=x^j  \;\;\;\;\;\; i\neq j=1,2
\eea
\ee
A solution to these equations is:
\be\bea{cc}
U_1=exp\{ -i R_1\bigl[a(p_1- {x^2 \over \pi{\cal F}})- {x^2 \over 
\pi{\cal F}}\bigr]\}\\
U_2=exp\{-i R_2\bigl[a(p_2+{x^1 \over {\cal F}})+ {x^1 \over 
{\cal F}}\bigr]\},
\eea\ee
with $a^2=1+{\pi^2{\cal F}^2\over R_1R_2}$. The above relations leads to
\be
U_1U_2 =e^{i\pi {\cal F}} U_2U_1.
\ee
This result is similar to the Matrix theory compactification on the NCT
formulated by CDS, described previously. It was argued there that, the 
noncommutativity of the torus is related to the non-vanishing of 3-form of 
M-theory, which in the string theory reduces to the antisymmetric NSNS 
2-form field, $B_{\mu\nu}$. In our case noncommutativity of the torus on
which the D-membrane of string theory is compactified, is a direct result of  
the non-vanishing $B$ field. In fact using the Matrix model formulation
of string theory [17], it is straightforward to obtain CDS results. In
the string matrix model of DVV, the matrices $X^{\mu}(\tau)$ of Matrix
theory, upon compactification of a space-like dimension, say $X^9$, become  
matrices $X^{\mu}(\tau,\sigma)$, receiving a $\sigma$ dependence, and
satisfying the Green-Schwarz action of the string theory in the light-cone frame; 
with their noncommutativity reflecting the added D-brane structure in string 
theory. Using the relation between conventional string theory and string 
matrix model, we map our noncommuting membrane coordinates $X^i,\;\ i=1,2$ 
, and the noncommutative torus generators $U_i$, to the Matrix theory 
and obtain  (2.4), (2.5) for compactification of Matrix theory 
on the NCT. 

The noncommutativity of the $c^*$ algebra (3.14) and (2.5) of the NCT is
similar to, but distinct from, the noncommutativity of the coordinates as in 
(3.10) and as it appears in Matrix theory and bound states of D-branes. 
The similarities are obvious, but the differences are subtle. In fact it is 
possible to see that when ${\cal F}$ is quantized to a rational number, by an
SL(2,Z) transformation, we can make the $U_1$ and $U_2$ commute, i.e. 
we can make the torus {\it commutative}, while the coordinates are 
{\it noncommutative}. 
Thus for {\it irrational} parameter $\theta$, we are dealing with a new 
form of noncommutativity not encountered in ordinary Matrix theory or
in the context of D-brane bound state.

\section{The BPS spectrum}
\setcounter{equation}{0} 
To complete our string theoretic description of the CDS formulation, i.e.
the SYM theory on NCT, we find the BPS spectrum of a system of (D2-D0)-brane 
bound state.
To have an intuitive picture, it is convenient to consider the T-dual version
of the mixed brane discussed earlier in section 3. The advantage of T-duality
is that in T-dual picture we only deal with commutative coordinates and 
commutative torus, where we are able to calculate the related spectrum by the
usual string theory methods.

Applying T-duality in an arbitrary direction, say $X^2$, (3.4) results in  
\be
\left\{  \begin{array}{cc}
\partial_{\sigma}X^0=0 \\
\partial_{\sigma}(X^{1}+{\cal F} X^2)=0 \\
\partial_{\tau}(X^{2}- {\cal F} X^{1})=0  \;\;\; \\
\partial_{\tau} X^a=0 \;\;\;\ ,\;\;\; a=3,...,9,
\end{array}\right.
\ee
describing a tilted D-string which makes an angle $\phi$ with the
duality direction, $X^2$: 
$$
\cot \phi={\cal F}.
$$
Thus we consider a D-string winding around a cycle of a torus defined by:
\be
\tau={R_2 \over R_1}e^{i\alpha}=\tau_1 +i \tau_2 \;\;\;\; , \;\;\;\;\
\rho=iR_1R_2\sin \alpha+b=i\rho_2+b,
\ee
where $b=BR_1R_2\sin\alpha$ is the flux of the $B$ field on the torus.
The D-string is located at an angle $\phi$ with the $R_1$ direction such that it 
winds $n$ times around $R_1$ and $m$ times around $R_2$. Hence
\be
\cot \phi={n \over m \tau_2}+\cot \alpha.
\ee
The greatest common divisor of $m$ and $n$ is the number of times D-string
winds around that cycle specified with the angle $\phi$.

The BPS spectrum of this tilted D-string system gets contributions from both
the open strings attached to the D-string and the D-string itself.
In order to consider the most general case, we assume a moving D-string  
which also has a non-zero electric field living on it.
\newline
{\it Open strings contributions} 

As in [23,24], the brane velocity and its electric field
will not affect the open strings spectrum. So it is sufficient to consider 
the open strings satisfying (4.1). These open strings have mode expansions [16]: 

\be\left\{ \bea{cc}
X^i=x_0^i + p^i \tau+ L^i \sigma + Oscil. \;\;\; , i=1,2  \\
X^0=x^0_0+p^0\tau+ Oscil. \\
X^a=x^a_0+ Oscil. \;\;\; , \;\;\; a=3,...,9 
\eea\right.
\ee
where $p^i$ and $L^i$, in usual complex notation, are:

\be 
p=\ r_1 {{n+m\tau} \over {|n+m\tau|^2}} \sqrt{{\tau_2\over \rho_2}}
\;\;\;\;\;\ ; r_1\in Z.
\ee
\be
L=\ q_1 {\rho{(n+m\tau)} \over {|n+m\tau|^2}} \sqrt{{\tau_2\over \rho_2}}
\;\;\;\;\;\ ; q_1\in Z.        
\ee
As we can see, $p$ is parallel to the D-string. We should note that, in the    
case of non-zero $B$ field, $L$ is no longer perpendicular to D-string. 
Moreover $|L|$ is an integer multiple of a minimum length. 
This is the length of a string stretched between two consecutive cycles of the 
wound D-string [25].

Mass of the open string defined by (4.4) is 
\be 
M^2=|p+L|^2+{\cal N}= {{\tau_2}\over {|n+m\tau|^2}}{{|r_1+q_1\rho|^2} \over 
\rho_2}+{\cal N},
\ee
where ${\cal N}$ is the contribution of the oscillatory modes. As it is seen,
(4.7) is manifestly invariant under both $SL(2,Z)$'s of the torus
acting on $\rho$ and $\tau$.
To find the contribution of the open string to BPS spectrum of the membrane
on $T^2_{\theta}$, we apply T-duality in $R_1$ direction,
$$
R_1 \rightarrow {1 \over R_1}\;\;\;\;\; {\rm or\ \ equivalently} 
\;\;\;\tau\leftrightarrow \rho,
$$
and obtain the spectrum of the open string compactified on NCT,

\be 
M^2={{\rho_2}\over {|n+m\rho|^2}}{{|r_1+q_1\tau|^2} \over \tau_2} 
+{\cal N},
\ee
with the $U(1)$ gauge field,
\be
{\cal F}^{-1}={n \over m\rho_2}+ \cot \alpha.
\ee
The above relation shows that ${\cal F}$ takes contributions from both the
torus ($\cot \alpha$) and the D-string tilt (${n \over m\rho_2}$). 
Rewriting (3.7) for T-dual of open string mode expansions (4.4), we get
\be
[X^1,X^2]=2\pi i\rho_2 ({n \over m}+\rho_2 \cot\alpha)^{-1}=2\pi i\rho_2(
{n\over m} -b)^{-1}.
\ee
The above equation is the same as the commutation relation between 
coordinates of a deformed torus in NCG, (2.14), where $\theta$ is substituted 
for the $b$ field, and the $b$ itself, through T-duality, is related to the 
angle of torus.

Note that a rational $b$ field will not give a NCT as shown in [4].
In our string theoretic description this is easily seen from the T-dual 
version, where, by means of a SL(2,Z) transformation we can transform such 
tori to an orthogonal torus, giving a zero $b$ field after T-duality.
Hence, only the irrational part of $b$ field can not be removed by $SL(2,Z)_N$
transformations.
\newline
{\it The D-string contribution}

Now we consider the most general case of a D-string on a torus, i.e. a moving 
D-string which has a non-zero electric field. To handle this problem, 
we use the DBI action which gives the dynamics of D-strings.
It has been shown in [23], for a D-string with an electric field
, [24] for the moving brane case, and [21], for a D-brane with a magnetic 
field, that the mass of a D-brane, calculated from DBI or string theory, coincide.

Consider the DBI action for the above tilted D-string moving with velocity $v$
normal to the D-string and the gauge field $F$ parallel to it in a non-zero  
$B_{12}$ background\footnote{ Because the DBI action for a $D_p$-brane has 
SO(p,1) symmetry, only the velocities normal to brane are relevant.}.
Here we assume that the $RR$ scalar is zero:
\be
S_{D-string}={-1\over g_s}\int d^2\sigma \sqrt{det(\eta_{ab}+{\cal F}_{ab})}.
\ee
For the D-string discussed above, we have
\be
\eta_{ab}=\left (\bea{cc}1-v^2 \;\;\;\;\ 0 \\ 0 \;\;\;\;\;\;\;\;\;\;  1  
\eea\right),
\ee
\be
{\cal F}_{ab}=\left (\bea{cc} 0 \;\;\;\;\ Bv+F \\ Bv+F \;\;\;\;\;\  0
\eea\right).
\ee
Inserting  (4.12), (4.13) in (4.11), we get 
\be
S_{D-string}={-1\over g_s}\int d^2\sigma \sqrt{1-v^2-(F+Bv)^2}.
\ee
To calculate the mass spectrum of D-string, we consider the Hamiltonian for
(4.14),
\be
H={1\over g_s}\int d\sigma \sqrt{1+[(P-\Pi B)^2+ \Pi^2]g_s^2},
\ee
where $P,\Pi$ are the conjugate momenta of collective coordinate of D-string 
and the electric gauge field, respectively: 
\be
P={\partial L \over \partial v}={1\over g_s}{v+B(F+Bv) \over 
\sqrt{1-v^2-(F+Bv)^2}}.
\ee
\be
\Pi={\partial L \over \partial F}={1\over g_s} {F+Bv \over \sqrt{1-v^2-(F+Bv)^2}}.
\ee
$P,\Pi$ defined on the dual torus, should be quantized [2]:
\be
P={r_2 \over \rho_2}{1 \over |n+m\tau|}  \;\;\;\; , \;\;\;\;\; 
\Pi={q_2 \over |n+m\tau|}.
\ee
Plugging (4.18) into (4.15), we have
\be
\alpha' M^2={|n+m\tau|^2 \rho_2 \over \alpha' g_s^2 \tau_2}+\alpha' {|r_2+\rho q_2|^2 \over
\rho_2 \tau_2}.
\ee
Applying T-duality on the (4.19), we obtain a (D2-D0)-brane bound state
in the presence of a non-vanishing $B$ field. Also we have turned on the 
electric fields living on the D2-branes world-volume. 
These electric fields are given by T-dual version of (4.18).
\newline
Hence the mass spectrum of the membrane discussed above is
\footnote{We should note that under T-duality the string coupling constant
behaves as $g_s\rightarrow g'_s=g_s\sqrt{{\tau_2 \over \rho_2}}$.}
\be
\alpha' M^2_{membrane}={|n+m\rho|^2 \tau_2 \over \alpha'  {g'_s}^2 \rho_2} 
+\alpha' {|r_2+\tau q_2|^2 \over\rho_2\tau_2}.
\ee
The $SL(2,Z)_N$ invariance, acting on $\rho$, is manifestly seen from the 
above equation.
\newpage
\noindent{\it The full spectrum}

As shown in [26], the open strings discussed earlier and the D-string 
form a {\it marginal} bound state, i.e. from the brane gauge theory point of 
view, the open strings are electrically charged particles with non-vanishing 
Higgs fields. So to find the full BPS spectrum, we should add the masses and 
not their square:
\be
{\cal M}= M_{membrane} + M_{open\  st.}.
\ee

\be
{\cal M}=\sqrt{{\tau_2 \over \rho_2}}{|n+m\rho| \over g'_s}
(1+{g'_s}^2{|r_2+q_2\tau|^2 \over \tau_2}{\rho_2 \over |n+m\rho|^2})^{1/2}+ 
{|r_1+q_1\tau|\over |n+m\rho|}\sqrt{{\rho_2 \over \tau_2}}.
\ee
The above spectrum is manifestly SL(2,Z) invariant. In the usual notation of 
$T^2_{\theta}$ [4], the $Sl(2,Z)_N$ acting on $\rho$ is non-classical. 

To compare our results with [4] or [19], we should find the zero volume and 
$g_s \rightarrow 0$ limits.
These limits are necessary for comparison, as the Matrix theory 
used in [4,19], is described by the 0-brane dynamics at small couplings, 
compactified on a light-like direction.

In the absence of a $B$ field background, applying the $SL(2,Z)_N$ 
transformation $\rho\rightarrow {-1 \over \rho}$, we can go to the 
large volume limit, sending the number of D0-branes to
infinity, reproducing the results of large $N$ Matrix model; but if the
$B$ field is non-zero the above transformation,
\be
iV+B\rightarrow {iV-B \over V^2+B^2},
\ee
does not allow $V$ to go to infinity. Hence the number 
of D0-branes distributed on D-membrane remains finite. 
In the zero volume limit, which is not altered by the above $SL(2,Z)_N$ 
transformation, we end up with a finite number of D0-branes.

In these limits  up to $V^2$ and $g_s$ 
$$
|n+m\rho|=|(n-m\theta)+iVm|=|n-m\theta|+{m^2V^2 \over 2|n-m\theta|}+O(V^3),
$$
and hence (4.22) reads as
\be
l_s{\cal M}={|n-m\theta| \over g_s} +{1 \over 2g_s}{m^2V^2 \over |n-m\theta|}
+{g_s \over 2|n-m\theta|}{|r_2+q_2\tau|^2 \over \tau_2}+
{|r_1+q_1\tau|\over |n-m\theta|}\sqrt{{V \over \tau_2}}.
\ee
The first term is due to the D2-brane itself 
\footnote{If we consider the full DBI action, which also has a constant 
term added due to D2-brane $RR$ charge, this term will be removed.}; 
the second term is due to the magnetic flux or the D0-brane contribution;
the third term is the contribution of electric fields; and the last term 
belongs to open strings.

We observe that this mass spectrum is equivalent to 
the spectrum given by Ho [19] (eq.(2.11)).
Comparing the spectrum (4.23) with the BPS spectrum of [4] and [19] we can 
construct the following correspondence table. 

\begin{table}[ht]
\caption{Comparison of parameters of [4],[19] with ours .\label{tab:sc}}
\vspace{0.3cm}
\begin{center}
\begin{tabular}{|c|c|c|c|}
\hline
$SYM/T^2_{\theta}$&$D=11\ SUGRA/S_- $&$ Matrix\ theory\ on T^2_{\theta}$&
$String\ theory\ (this\ work)$ \\
 \hline
$P^i=\int T^{0i}$&$nm^{i-}$&$n\epsilon^{ij}m_{j-}=nm^i$& a combination \\
$e^i$&$e^i$&$n^i$&of $ r_2,\ q_2$\\
$p'^i$&$w^i$&$w^i$&$ r_1,\ q_1$\\
$p$&$n\ (p_-={n-m\theta \over R})$ &$n$&$n$\\
$q$&$m^{ij}$&$m=m^{ij}\epsilon^{ij}$&$m$\\
$\theta$&$RC_{-ij}$&$\theta$&$b=B\ field\ flux$\\ 
\hline
\end{tabular}
\end{center}
\end{table}

\noindent
As we see the interpretation of $p_-={n-m\theta \over R}$ in our case is 
the mass of the tilted D-string which is closely 
related to the DVV's string matrix theory in the noncommutative case. 

The $SL(2,Z)_N$ symmetry generators of (4.22) are 
$$
\rho \rightarrow \rho+1 \;\;\;\;\; ,\;\;\;\;\;\; \rho\rightarrow {-1\over \rho}
$$
which in the zero volume limit ($\rho_2=0$) become
\be
\theta\rightarrow \theta+1 \;\;\;\ , \;\;\; \theta\rightarrow {-1\over \theta}
\ee
Under the above transformations $(n,m)$ transforms as an $SL(2,Z)_N$ vector.
There is also an $SL(2,Z)_C$ symmetry of mass spectrum (4.23), acting on the 
$\tau$, under which the $(r_i,q_i)$ behave as SL(2,Z) vectors. 

Invariance of the mass spectrum, (4.23), under $\theta\rightarrow 
{-1\over \theta}$, implies that
\be
g_s\rightarrow g'_s=g_s {\theta}^{-1} 
\ee
Moreover the imaginary part of $\rho\rightarrow {-1\over \rho}$, tells us that 
the volume of the torus in the zero volume limit, in the string theory units, 
transforms as:
\be
V\rightarrow V'=V \theta^{-2}
\ee
Putting these relations together, and remembering the relation of 
10 dimensional units and 11 dinemsional parameters, $l_p^3=l_s^3 g_s$ 
and $l_sg_s=R$, and assuming $l_p$  invariance under $\theta$  
transformations, we obtain: 
\be\bea{cc}
R \rightarrow R'=R {\theta}^{-2/3}  \\
R_i\rightarrow R'_i=R_i {\theta}^{-2/3} \\
l_s\rightarrow l'_s=l_s {\theta}^{-1/3}. 
\eea\ee
The above relations differ from the corresponding relation [4] or [19] 
(eq.(2.13)) and indicate an M-theoretic origin for the $SL(2,Z)_N$. 
This is the effect of considering the whole DBI action and not, 
only its second order terms. 
The same result is also obtained from DBI by Ho [27].

\section{ Discussion}
\setcounter{equation}{0} 
In this work we have studied more extensively the brane systems in a non-zero 
$B_{\mu\nu}$ background field, through the usual string theory methods, a
problem also considered in [13,14,15].
\newline
In [14] D0-brane dynamics in $B$ field background was considered,
and shown that the $B$ field modifies the D0-brane dynamics. As 
discussed there, the effects of such a background is to replace the Poisson bracket
of fields by the noncommutative version, the Moyal bracket. The key idea 
there, is that the existence of background $B$ field introduces a 
{\it phase factor} for the open strings attached to D0-branes, the phase 
factor being proportional to the background $B$ field. These open strings, as 
discussed in [2,3] carry the dynamical degrees of
freedom of D-branes. Using this phase, they showed that this background 
will lead to a NC background in the related Matrix model.  
The same procedure in a slightly different point of view was considered in [15], 
supporting and clarifying the novel result in[4]. 

In this paper, extending the string theoretic ideas of [14,15,16], we  have
explicitly shown that the noncommutativity of brane coordinates come about 
naturally in the formulation of open strings in the background $B$ field,
as well as the noncommutativity of the torus. 

To use the usual string theory methods, by means of T-duality, we replaced
the torus defined by $\tau={R_2 \over R_1}e^{i\alpha}$ , 
by a torus with a $B$ field, where the $B$ flux is $R_1R_2\cos \alpha$.
Hence in the
T-dual version we dealt with a D-string wound around the cycle of the torus.
Using the usual string theoretic arguments and also the DBI action for 
the corresponding D-string dynamics, we calculated the BPS spectrum of a 
system of (D2-D0)-brane bound state, in a $B$ field background. 
As shown here, this brane system is described by a DBI action formulated on
a noncommutative torus. 
In a remarkable paper [13] Li argued that SYM on $T^2_{\theta}$
will not fully describe the dynamics of D0-branes in a background $B$ field,
and DBI action becomes necessary.
Our spectrum, in the small coupling and zero volume
limit, reproduces the CDS results. 

A novel feature in our work is that under the transformation  
$\rho\rightarrow {-1 \over \rho}$, $R_i$ and the 
eleventh dimension compactification radius, $R$, transform in the same way; 
$R,R_i\rightarrow R',R'_i=R,R_i {\theta}^{-2/3}$,
in contrast to the results of [4,19] where 
SYM and  Matrix model rather than the DBI action on a noncommutative 
torus, were used.

It is amusing that the dimension of ${\cal H}$ [4], $dim{\cal H}=|n-m\theta|$,
is given by the length of the wrapped D-string. 
$dim {\cal H}=Tr{\bf 1}=|n-m\theta|$ is a factor coming in front of the YM 
action on $T^2_{\theta}$, which up to second order of ${m\over n}$ reproduces the
higher power terms of DBI action in a $B$ field background,  as discussed in
[13].

There are still a number of questions to be addressed: One is, how to
realize the $IIB$ SL(2,Z) in the Matrix theory on NCT.
In other words, if we have both $B_{\mu\nu}$ and $\tilde{B_{\mu\nu}}$ (the  
RR two form), as background fields, how to incorporate 
this in the Matrix model on NCT. As indicated by Ho [28]
, the $\tilde{B_{\mu\nu}}$ is related to the
$g_{-i}$ component of the metric. So the problem of $g_{-i}$ addressed by CDS, 
seems to be related to $\tilde{B}$ background.

Another interesting open question briefly studied in [11,29] is in relation 
to the six dimensional theories. 
Six dimensional theories as theories living in the NS5-brane world 
volume, show up in the compactification of Matrix model on $T^n, \;\ n>4$.
These theories seem to be non-local field theories.
On the other hand along the arguments of [10,11], we know that considering 
$B$ field background leads to non-local low energy field theories for open 
strings. The method we used here, i.e. applying T-duality to remove 
noncommutativity and replacing $B$ field with the torus angle, may give new 
insight into the problem of 
NS5-branes (or six dimensional theories) in ordinary string theory. 

{\bf Acknowledgements}

We would like to thank P. Ho  and A. Fatollahi for helpful comments.

\end{document}